\documentclass[12pt]{article}

\usepackage{graphicx}

\begin{document}
\begin{center}
{\Large\bf \boldmath How to extract information \\[2mm]
               from Green's functions in Landau gauge} 

\vspace*{6mm}
{Attilio Cucchieri$^a$ and Tereza Mendes$^{a,b}$ }\\      
{\small \it $^a$ Instituto de F\'\i sica de S\~ao Carlos, Universidade de S\~ao Paulo, \\
                 Caixa Postal 369, 13560-970 S\~ao Carlos, SP, Brazil \\   
            $^b$ DESY, Platanenallee 6, 15738 Zeuthen, Germany
            }
\end{center}

\vspace*{6mm}

\begin{abstract}
The infrared behavior of gluon and ghost propagators offers a crucial test of
confinement scenarios in Yang-Mills theories. A nonperturbative study of
these propagators from first principles is possible in lattice simulations,
but one must consider significantly large lattice sizes in order to approach
the infrared limit. We propose constraints based on general properties of the
propagators to gain control over the extrapolation of data to the infinite-volume
limit. These bounds also provide a way to relate the propagators to simpler,
more intuitive quantities. We apply our analysis to the case of pure $SU(2)$ gauge
theory in Landau gauge, using the largest lattice sizes to date. Our results seem
to contradict commonly accepted confinement scenarios. We argue that it is not so.
\end{abstract}


\vspace*{6mm}

\section{The Gribov-Zwanziger Confinement Scenario}
\label{sec:GZ}

About thirty years ago, Gribov proposed an interesting confinement mechanism
for color charges in Landau (and Coulomb) gauge \cite{Gribov:1977wm}.
His idea was based on the restriction of the physical configuration space
to the region $\Omega$ of transverse configurations, delimited by the
so-called first Gribov horizon, where the smallest (non-trivial) eigenvalue
of the Faddeev-Popov (FP) operator ${\cal M} = - D_{\mu} \partial_{\mu}$ is
zero. The limitation of the functional integration to the (first Gribov)
region $\Omega$ was an attempt to fix the gauge completely, getting rid of
spurious gauge copies, known thereafter as Gribov copies. Since the ghost
propagator $G(p)$ is given by $\langle p | {\cal M}^{- 1} | p \rangle$ and
${\cal M}$ is semi-positive definite for gluon fields $A \in \Omega$, one
cannot have a singularity for $G(p)$ at a finite momentum $p$. Using
perturbation theory up to second order, Gribov wrote (for Landau gauge)
the no-pole condition \cite{Gribov:1977wm,Sobreiro:2005ec}
\begin{equation}
\sigma(0) \, = \, \frac{N_c}{4 \left(N_c^2 -1 \right)}
  \, \int \frac{d^4q}{(2 \pi)^4} \frac{\langle \, A^{a}_{\lambda}(q)
                                       \, A^{a}_{\lambda}(-q) \, \rangle}{q^2}
                 \, < \, 1 \; .
\label{eq:sigma0}
\end{equation}
Here $N_c$ refers to the gauge group $SU(N_c)$, an average over the
Lorentz indices $\lambda$ has been considered and the quantity $\sigma(p)$
enters the ghost propagator as $ G(p) \approx p^{-2} \left[ 1 - \sigma(p)
\right]^{-1} $. The above inequality tells us that, in the infrared (IR) limit,
the Landau gluon propagator $ D^{ab}_{\mu \nu}(p) = \langle A^{a}_{\mu}(p)
A^{b}_{\nu}(-p) \rangle$ is less singular than $1/p^{2}$ (in the $4d$ case).
Moreover, by using the above no-pole condition as a characterization of
the first Gribov region, one can show that the tree-level gluon propagator
becomes
\begin{equation}
D(p) \, = \, g_0^2 \, p^2 / (p^4 + \gamma^4) \; ,
\label{eq:dgribov}
\end{equation}
where the (Gribov) mass parameter $\gamma$ is fixed by the gap equation
$ \int d^4q (2 \pi)^{-4} (q^4 + \gamma^4)^{-1} = 4/(3 N_c g^2) $. Then,
by using (\ref{eq:dgribov}) and the
gap equation, one can study the IR limit of $\sigma(p)$, obtaining an
IR-enhanced ghost propagator $G(p) \propto 1/p^4$. As stressed by Gribov
\cite{Gribov:1977wm}, this enhancement is an indication of a long-range
effect in the theory that may explain color confinement. In Coulomb gauge,
this enhancement of $G(p)$ can be directly related to a color-Coulomb
potential linearly rising with distance.

It is interesting that a bound similar to the one
above has also been obtained in Ref.\ \cite{Dell'Antonio:1989jn} by
considering a variational method applied to the FP operator ${\cal M}$.
This bound, known as the ellipsoidal bound, can be written (in the continuum) as
\begin{equation}
\int \frac{d^dq}{(2 \pi)^d} \frac{\langle A^{b}_{\mu}(q)
                                     \, A^{b}_{\mu}(-q) \rangle}{q^2}
                       \, \leq \, C \; ,
\label{eq:ellbound}
\end{equation}
where $C$ depends on the dimensionality $d$ of the space-time and
on the gauge group. One should stress that the ellipsoid ${\cal E}$, defined by
the ellipsoidal bound, is a region of transverse configurations that includes the
first Gribov region $\Omega$. A similar bound can also be obtained on the lattice
\cite{Zwanziger:1991gz}. Moreover, it is convenient to define on the lattice a
region $\Theta$, included in the ellipsoid ${\cal E}$ and including
the first Gribov region $\Omega$, i.e.\ $\Omega \subset \Theta \subset {\cal E}$. For all
configurations belonging to $\Theta$, one can prove \cite{Zwanziger:1990by}
\begin{equation}
\left| \, {\widetilde A}^b_{\mu}(0) \, \right| \, \leq \,
               2\tan{\left(\frac{\pi}{V^{1/d}}\right)} \; ,
\label{eq:M0L}
\end{equation}
where $V$ is the lattice volume and $ {\widetilde A}^b_{\mu}(0) = V^{-1} \sum_x
A^b_{\mu}(x) $ is the gluon field at zero momentum. This quantity may be viewed
as a magnetization $M$. Thus, in the infinite-volume limit one can show that
$|M|$ is zero. By adding an external color ``magnetic'' field $H$ coupled to
$A$ in the action and using the above inequality, one also obtains that the
free energy per unit volume is null when $V$ goes to infinity \cite{Zwanziger:1990by}.
If $H$ is spatially modulated, then the susceptibility $\chi$ at zero external
field coincides with the usual gluon propagator $D(p)$. The inequalities valid in
the region $\Theta$ would then suggest that $\lim_{p \to 0} \chi(H=0,p)
= \lim_{p \to 0} D(p) = 0$ \cite{Zwanziger:1991gz}.

In order to restrict the functional integration to the first Gribov region
$\Omega$, Zwanziger added to the usual Yang-Mills action a non-local term
proportional to $ {\cal M}^{- 1} $ \cite{Zwanziger:1989mf}. This term clearly
suppresses the probability of configurations near the boundary $\partial \Omega$ of the
region $\Omega$. After localizing the action, zeroth-order perturbation theory allows one
to obtain again the propagator in Eq.\ (\ref{eq:dgribov}). The purely imaginary poles
$p^2 = \pm i \gamma^2$ of this propagator make it incompatible with a Kallen-Lehmann
representation \cite{Zwanziger:1989mf}. These singularities at an unphysical
location suggest that gluons are indeed not physical excitations. One should
also note \cite{Zwanziger:1991gz,Zwanziger:1990by} that $ D(0) = \int d^dx \, D(x) = 0 $
means a gluon propagator in position space $D(x)$ that is positive and negative in equal
measure. This represents a maximal violation of reflection positivity. In Refs.\
\cite{Zwanziger:1991gz,Zwanziger:1990by} the violation of reflection positivity
for the gluon propagator has been proposed as a confinement mechanism for
gluons.

The restriction of the functional integration to $\Omega$ has also been
discussed in Ref.\ \cite{Zwanziger:1991ac}. The non-local term
$\, - \gamma^4 {\cal H} \,$ is added to the Yang-Mills action, where ${\cal H}$
is the so-called horizon function, containing an $ {\cal M}^{- 1} $ factor. At the same time, the
Gribov mass $\gamma$ is fixed implicitly by the horizon condition $ \langle h \rangle
= (N_c^2 - 1) d $, where $h$ is the horizon function per unit volume. It is interesting
that the horizon condition implies a ghost propagator enhanced in the IR limit
\cite{Zwanziger:2001kw}, i.e.\ $\lim_{p\to0} [ p^2 G(p) ]^{-1} = 0 \,$. Clearly,
the enhancement of the ghost propagator at $p = 0$ should indicate that $G(p)$ feels
the singularity of $ {\cal M}^{- 1} $ on $\partial \Omega$. Indeed, since the
configuration space has very large dimensionality one expects that in the
infinite-volume limit, due to entropy considerations, the Boltzmann weight be
concentrated on $\partial \Omega$ \cite{Zwanziger:1991ac}. This implies that the smallest
nonzero eigenvalue $\lambda_{min}$ of ${\cal M}$ should go to zero in the infinite-volume
limit. (This has been verified numerically in Landau gauge \cite{lambdamin}.)


\vskip 3mm

The IR behavior of propagators and vertices in Landau gauge has also been studied
in Ref.\ \cite{Alkofer:2008jy} by considering the sets of coupled
Dyson-Schwinger equations (DSE) for the basic propagators and vertices of
Yang-Mills theory (in the $4d$ case). By using a simple power counting for
the solutions in the IR limit and constraints obtained using a skeleton expansion,
the authors found a consistent solution characterized by an IR-enhanced ghost
propagator $G(p) \sim p^{-2(1+\kappa)}$ and by an IR-finite gluon propagator
$D(p) \sim p^{2(2\kappa-1)}$ with $\kappa \in [1/2,3/4]$. Note that $D(0) = 0$
for $\kappa > 1/2$. On the other hand, the analysis carried out in
\cite{Alkofer:2008jy} allows also for a solution with a
tree-level-like ghost propagator at small momenta $G(p) \sim p^{-2}$ and a
finite nonzero gluon propagator $D(0) > 0$. These two consistent solutions
have also been obtained by several studies of DSE using specific
approximations \cite{DSE}.

Recently it was shown \cite{Dudal:2007cw,3d4d} that using
the Gribov-Zwanziger approach, i.e.\ by restricting the functional integration
to the Gribov region $\Omega$, one can also obtain in $3d$ and $4d$ a finite nonzero
gluon propagator $D(0)$ and a tree-level-like ghost propagator in the IR limit.
Here the dynamical mechanism is related to a suitable mass
term that may be added to the action while preserving its renormalizability.
As a consequence, one can show that the restriction to $\Omega$
induces a soft breaking of the BRST symmetry \cite{3d4d}.
It is interesting that the same approach cannot be extended to
the $2d$ case \cite{Dudal:2008xd}, because the new mass term produces IR
singularities that make the restriction to $\Omega$ impossible.

An IR-enhanced Landau ghost propagator is also obtained as a consequence of
the so-called confinement criterion of Kugo-Ojima \cite{Kugo:1979gm}. At the
same time, this criterion suggests that the perturbative massless pole in the transverse
gluon propagator should disappear \cite{Kugo:1995km}. In this sense, an
IR-suppressed gluon propagator (not necessarily vanishing) can be accommodated
in this confinement scenario \cite{Alkofer:2000wg}. Finally, even though the
Gribov-Zwanziger and the Kugo-Ojima confinement scenarios seem to predict similar
IR behavior for
the propagators, it is not clear how to relate the (Euclidean) cutoff at
the Gribov horizon to the (Minkowskian) approach of Kugo-Ojima \cite{Zwanziger:2003cf}.


\section{Bounds for the Gluon and the Ghost Propagators}

Recently we have introduced \cite{Cucchieri:2007rg} rigorous upper
and lower bounds for the  gluon propagator at zero momentum $D(0)$
by considering the quantity
\begin{equation}
{M}(0) \, = \, \frac{1}{d (N_c^2 - 1)}
   \sum_{b,\mu} | {\widetilde A}^b_{\mu}(0) | \; ,
\end{equation}
with $ {\widetilde A}^b_{\mu}(0) $ defined in the previous section. Indeed,
by straightforward calculations one finds that
\begin{equation}
V \, {\langle {M}(0) \rangle}^2 \, \leq \; D(0)\;
    \leq \; V d (N_c^2 - 1) \, \langle {{M}(0)}^2 \rangle \; .
\label{eq:Dbounds}
\end{equation}
Thus, if ${M}(0)$ goes to zero as $V^{-\alpha}$ we obtain that $ D(0)
\to 0$, $0 < D(0) < +\infty$, or $ D(0) \to +\infty $, respectively if
$\alpha$ is larger than, equal to or smaller than 1/2. Recall that
${M}(0)$ should go to zero at least as $V^{-1/d}$ in the $d$-dimensional
case [see Eq.\ (\ref{eq:M0L})]. At the same time, a necessary condition
to find $D(0)=0$ is that ${M}(0)^2$ goes to zero faster than $1/V$. We
note that the above bounds apply to any gauge and that they can be
immediately extended to the case $D(p)$ with $p \neq 0$.

We investigated the bounds (\ref{eq:Dbounds}) for pure $SU(2)$ gauge theory in
Landau gauge considering several lattice volumes in $2d$, in $3d$ and in
$4d$ with the largest lattice corresponding (respectively) to $a^2 V \approx
(70 \,\mbox{fm})^2$, $a^3 V \approx (85 \, \mbox{fm})^3$ and to $a^4 V
\approx (27 \,\mbox{fm})^4$. By using the Ansatz $B_u / L^u$ for $ a^2
\langle \, {M}(0)^2 \, \rangle $ we obtain $u = 2.72(1)$ in the $2d$ case,
implying $D(0)=0$. A similar analysis in $3d$ and in $4d$ for the lower and
the upper bounds gives $0.4(1)$ GeV$^{-2} \leq a^2 D(0) \leq 4(1)$ GeV$^{-2}$
in $3d$ and $2.2(3)$ GeV$^{-2} \leq a^2 D(0) \leq 29(5)$ GeV$^{-2}$ in $4d$.
Recently, a study for the $4d$ $SU(3)$ case \cite{Oliveira:2008uf} also
finds a value for $\alpha$ very close to $1/2$. Although the authors conclude that
$D(0)=0$ in the infinite-volume limit, one should observe that in this case
the lattice volumes considered are relatively small and the statistics
is rather low. Thus, a more detailed analysis in the
$SU(3)$ case should be carried out in order to verify if the IR behavior of
the gluon propagator agrees \cite{Cucchieri:2007zm} or not with the $SU(2)$ case.

One can also obtain lower and upper bounds for the ghost propagator
\cite{Cucchieri:2008fc}. In Landau gauge, for any nonzero momentum $p$, one
finds
\begin{equation}
\frac{1}{N_c^2 - 1} \,  \frac{1}{\lambda_{min}} \, \sum_a \,
  | {\widetilde \psi_{min}(a,p)} |^2 \,
              \leq \, G(p) \, \leq \, \frac{1}{\lambda_{min}} \; ,
\label{eq:Gineq}
\end{equation}
where $\lambda_{min}$ is the smallest nonzero eigenvalue of the FP
operator ${\cal M}$ and ${\widetilde \psi_{min}(a,p)}$ is the corresponding eigenvector.
If we assume $\lambda_{min}\sim N^{-\delta}$ and
$G(p) \sim p^{-2-2\kappa}$ at small $p$, we should find
$ 2+2\kappa \leq \delta$, i.e.\ $\delta > 2$ is a necessary condition for
the IR enhancement of $G(p)$. Note that a similar analysis can be carried out 
\cite{Cucchieri:2006hi} for any generic gauge condition ${\cal F}[A]=0$
imposed on the lattice by minimizing a functional $E[U]$, where
$U$ is the (gauge) link variable. Indeed, from the second variation of $E[U]$
one can obtain the corresponding FP matrix ${\cal M}$ and the set of local
minima of $E[U]$ defines the Gribov region $\Omega$, where all eigenvalues
of ${\cal M}$ are positive. In the infinite-volume limit,
entropy favors configurations near $\partial \Omega$
(where $\lambda_{min}$ goes to zero). Thus, inequalities of the type (\ref{eq:Gineq})
can tell us if one should expect an enhancement of the ghost propagator $G(p)$ 
when the Boltzmann weight gets concentrated on $\partial \Omega \,$.\footnote{For
example, in $4d$ Maximally Abelian gauge one sees
that $\lambda_{min}$ goes to zero at large volume but the ghost propagator
stays finite at zero momentum \cite{MAG}.}

A study in the $SU(2)$ Landau case \cite{Cucchieri:2008fc} suggests that $\delta>2$
in 2d, implying IR enhancement of $G(p)$, while $\delta<2$ in 4d. These results are
confirmed if one considers the dressing function $p^2 G(p^2)$ for very large lattice
volumes \cite{Cucchieri:2008fc}. Indeed, the data in the $2d$ case can be fitted by
$\sim p^{-2 \kappa}$, with $\kappa$ between 0.1 and 0.2. On the contrary, in $3d$ and
in $4d$ the data are well described by $a - b \log(1 + c p^2)$, supporting $\kappa
= 0$.

Let us note that our data for the gluon and ghost propagators are in good
agreement with results obtained by other groups using very large lattice
volumes \cite{largevolume}. Of course, one should also recall that the region $\Omega$
is actually not free of Gribov copies \cite{Dell'Antonio:1989jn,Zwanziger:1990by,
Dell'Antonio:1991xt} and that the configuration space should be identified with
the so-called fundamental modular region (FMR) $\Lambda$. On the other hand,
the restriction of the configuration space to the FMR should not make any
difference on the numerical verification of the Gribov-Zwanziger scenario.
Indeed, as we have seen in the previous section, this scenario is based on the
restriction of the configuration space to the region $\Omega$, which includes
$\Lambda$. Actually, the bounds obtained for the gluon fields (see again Section
\ref{sec:GZ}) apply to regions, such as $\Theta$ and ${\cal E}$, that are even
larger than the region $\Omega$. Finally, as explained in \cite{Cucchieri:1997dx},
the restriction to the FMR can only make the ghost propagator less singular, as
confirmed by recent lattice data \cite{gribovghost}.


\section{Conclusions}

We have presented simple properties of gluon and ghost propagators that constrain
(by upper and lower bounds) their IR behavior. For the gluon case we define a
magnetization-like quantity, while for the ghost case we relate the propagator to
$\lambda_{min}$ of the FP matrix. We propose the study of these quantities, as a
function of the lattice volume, in order to gain better control of the infinite-volume
limit for the propagators in the IR regime.\footnote{Note that a faster approach to the
infinite volume could be obtained by using an extended gauge-fixing as done in
\cite{Bogolubsky:2007bw}.}

Our data support a Landau-gauge gluon propagator that is IR finite in $3d$ and $4d$. This
result can be interpreted \cite{Cucchieri:2007rg} as a consequence of ``self-averaging''
of a magnetization-like quantity, i.e.\ $M(0)$ without the absolute value. In particular,
one may think of $D(0)$ as a response function (susceptibility) of this magnetization.
In this case it is natural to expect $D(0) > 0$ in the infinite-volume
limit.\footnote{The massive behavior displayed by the gluon propagator in the IR limit has
been recently criticized \cite{AS} due to the observation that different lattice
discretizations yield different mass values in lattice units. On the other hand, such a
comparison only makes sense when data are extrapolated to the continuum limit and that,
of course, is not the case when the simulations are done at $\beta = 0$.} In the
$2d$ case the magnetization is ``over self-averaging'' and the susceptibility is zero.
These results are in agreement with the suppression of the IR components of the
gluon field $A$ due to the limitation of the functional space to the first Gribov region
$\Omega$. At the same time the gluon propagator displays a clear violation of reflection
positivity in Landau gauge \cite{violation}, i.e.\ the confinement mechanism for
gluons proposed in \cite{Zwanziger:1991gz,Zwanziger:1990by} is confirmed by lattice data.

For the ghost propagator we find that in $3d$ and $4d$ the behavior at small momenta is essentially
tree-level like, while in $2d$ this propagator seems to be clearly enhanced compared to
the perturbative behavior $p^{-2}$. As described in Section \ref{sec:GZ}, these
results are not necessarily in contradiction with the Gribov-Zwanziger approach
\cite{Dudal:2007cw,3d4d,Dudal:2008xd}.


\section{Acknowledgments}

We acknowledge partial support from FAPESP and from CNPq. The work of T.M. was
supported also by the Alexander von Humboldt Foundation.



\end{document}